\documentclass[nonacm,sigconf]{acmart}

\AtBeginDocument{%
  }

\acmConference[]{}{}{}

\usepackage{algorithm}
\usepackage{algpseudocode}
\usepackage{pgfplots}
\usepackage{siunitx}
\usepackage{multirow}
\usepackage{booktabs}
\usepackage{cleveref}
\usepackage{subcaption}
\pgfplotsset{compat=1.18}

\usepackage{pifont}
\newcommand{\cmark}{\ding{51}}
\newcommand{\xmark}{\ding{55}}

\definecolor{RoyalBlue}{RGB}{65,105,225}
\definecolor{OliveGreen}{RGB}{85,107,47}
\definecolor{Magenta}{RGB}{255,0,255}
\definecolor{Apricot}{RGB}{251,206,177}
\definecolor{RedOrange}{RGB}{255,83,73}

\author{Suhrid Gupta}
\email{suhrid.gupta1@unimelb.edu.au}
\affiliation{%
  \institution{The University of Melbourne}
  \city{Melbourne}
  \state{Victoria}
  \country{Australia}
}

\author{Muhammed Tawfiqul Islam}
\email{tawfiqul.islam@unimelb.edu.au}
\affiliation{%
  \institution{The University of Melbourne}
  \city{Melbourne}
  \state{Victoria}
  \country{Australia}
}

\author{Rajkumar Buyya}
\email{rbuyya@unimelb.edu.au}
\affiliation{%
  \institution{The University of Melbourne}
  \city{Melbourne}
  \state{Victoria}
  \country{Australia}
}

\begin{document}

\title[]{A Hybrid Reactive-Proactive Auto-scaling Algorithm for SLA-Constrained Edge Computing}

\begin{abstract}
Edge computing decentralizes computing resources, allowing for novel applications in domains such as the Internet of Things (IoT) in healthcare and agriculture by reducing latency and improving performance. This decentralization is achieved through the implementation of microservice architectures, which require low latencies to meet stringent service level agreements (SLA) such as performance, reliability, and availability metrics. While cloud computing offers the large data storage and computation resources necessary to handle peak demands, a hybrid cloud and edge environment is required to ensure SLA compliance. This is achieved by sophisticated orchestration strategies such as Kubernetes, which help facilitate resource management. The orchestration strategies alone do not guarantee SLA adherence due to the inherent delay of scaling resources. Existing auto-scaling algorithms have been proposed to address these challenges, but they suffer from performance issues and configuration complexity. In this paper, a novel auto-scaling algorithm is proposed for SLA-constrained edge computing applications. This approach combines a Machine Learning (ML) based proactive auto-scaling algorithm, capable of predicting incoming resource requests to forecast demand, with a reactive autoscaler which considers current resource utilization and SLA constraints for immediate adjustments. The algorithm is integrated into Kubernetes as an extension and its performance is evaluated through extensive experiments in an edge environment with real applications. The results demonstrate that existing solutions have an SLA violation rate of up to 23\%, whereas the proposed hybrid solution outperforms the baselines with an SLA violation rate of only 6\%, ensuring stable SLA compliance across various applications.
\end{abstract}



\maketitle

\section{Introduction}
\label{sec:introduction}

Cloud computing architectures leverage the on-demand accessibility of the Internet. The applications deployed here utilize the vast resources of the cloud to perform a task and relinquish it once it is complete for the other sub-modules in the application to request~\cite{rimal2009taxonomy}. In the early days, a singular end-point would be used to access these services, however nowadays the architecture is a multi-regional model allowing effortless access from across the world. The increasing popularity of hand-held devices as well as home appliances has resulted in data being largely produced at the edge of the cloud network. Thus, processing this large amount of data solely on the cloud proved to be an inefficient solution due to the bandwidth limitations of the network~\cite{shi2016edge} This was resolved through the use of edge computing architectures.

Edge computing ensures data processing services and resources exist at the peripheries of the network~\cite{cao2020overview}. The architecture extends and adapts the computing and networking capabilities of the cloud to meet real-time, low latency, and high bandwidth requirements of modern agile businesses.

Edge computing deploys several lightweight computing devices known as \textit{cloudlets} to form a ``mini-cloud'' and places them in close proximity to the end-user data~\cite{liu2019survey}. This reduces the latency in terms of client-server communication and data processing. Cloudlets can also be easily scaled depending on the resource requirements per edge architecture. However, due to the dynamic resource requirements which may fluctuate from time to time, the resources allocated to cloudlets must be dynamically scaled too. This dynamic scaling, along with the inherent latency present between the cloud layer and the edge cloudlets, poses a significant problem to real-time resource scaling~\cite{varghese2016challenges}.

One method of mitigating this scaling latency is through the use of microservice applications. By employing a microservice architecture, the resources in a cloudlet are distributed as a collection of smaller deployments that are both independent and loosely coupled~\cite{villamizar2015evaluating}. This loose coupling ensures that parts of the cloudlet can be scaled as required, further reducing the time required to scale resources as compared to scaling the cloudlet monolithically.

The scaling of these microservice resources is done automatically through a process known as auto-scaling. While most container orchestration platforms come bundled with default auto-scaling solutions, and these solutions are sufficient for most applications, they fall apart when scaling resources for time-sensitive services processing real-time data. Applications such as the ones used in healthcare require stringent compliance to service level agreements (SLA) on metrics such as application latency. This has led to further research on auto-scaling solutions for edge computing applications. These primarily fall into two categories:

\textbf{Reactive auto-scaling solutions} attempt to modify the microservice resource allocation once the required resources exceed the current allocation. These algorithms are simple to develop and deploy, however, the time taken to scale resources leads to a degradation of resource availability and violates SLA compliance~\cite{podolskiy2018iaas}.

\textbf{Proactive auto-scaling solutions} attempt to model resource allocation over time and effectively predict the resource requirements. By doing so, the microservice resources can be scaled in advance through a process known as ``cold starting''. This approach removes the latency inherent in scaling resources, however, the algorithms are extremely complex to develop, train, and tune to specific edge applications~\cite{straesser2022not}.

To tackle these challenges, this paper proposes a hybrid approach that combines the simplicity of using reactive autoscalers, while maintaining the resource availability benefits of the proactive autoscalers. The algorithm involves a smaller-scale LSTM machine learning model which can be quickly trained to recognize the key features of the resource workload over a course of time. The autoscaler then uses the predictions from the LSTM model to scale its resources in advance. A reactive autoscaler is used to maintain the current resource requirements as the predictive model cannot provide fine-tuned predictions. The accuracy of the prediction model is gauged by monitoring the SLA metrics. If it is observed that the predictive model is performing poorly, the training parameters are automatically tuned for the next training iteration.

The \textbf{contributions} of this paper are as follows:
\begin{itemize}
    \item Propose a hybrid auto-scaling method that mitigates the challenges present in reactive and proactive methods.
    \item The algorithm is implemented as an extension to Kubernetes.
    \item Deploy this algorithm in a real edge cluster prototype.
    \item Conduct extensive experiments using realistic daily workloads on a prototype microservice application.
    \item Compare the SLA violations of cutting-edge reactive, proactive, and default container orchestration autoscalers with the proposed hybrid algorithm.
\end{itemize}

The remainder of the paper is organized as follows. Section~\ref{sec:related-work} discusses related autoscaling strategies, their benefits, and drawbacks. Section~\ref{sec:problem-formulation} formulates the problem we are attempting to solve, including SLA and the cold start problem. Section~\ref{sec:hybrid-design} provides an overview of the new hybrid autoscaler architecture and algorithm along with an analysis of its computational complexity. Section~\ref{sec:experimental-setup} deals with the experiments conducted to validate the efficacy of the proposed architecture, with Section~\ref{sec:evaluation} discussing the performance of the architecture. Finally, Section~\ref{sec:conclusion} concludes and summarizes the findings and discusses some of the future enhancements that could be made to the architecture.

\section{Related Work}
\label{sec:related-work}

\subsection{Reactive Autoscaling Strategies}
\label{subsec:reactive}

Nunes et al.~\cite{nunes2021state} stated that horizontal pod auto-scaling using a reactive strategy remains the most popular auto-scaling technique, as well as research topic. These strategies, despite having limitations such as a reliance on predetermined resource thresholds and a delay in resource scaling, have been popular in research articles. Dogani et al.~\cite{dogani2023auto} stated that this was due to the simplicity and user-friendliness in developing them.

Kampars and Pinka~\cite{kampars2017auto} proposed a reactive auto-scaling algorithm for edge architectures based on open-source technologies. The algorithm scales in a non-standard approach, considering real-time adjustments in the application logic to determine the strategy of scaling. Zhang et al.~\cite{zhang2019quantifying} presented an algorithm for determining edge elasticity through container-based auto-scaling, demonstrating that balancing system stability with decent elasticity required careful tuning of parameters such as cooldown periods. However the lack of addressing the cold start problem results in a delay in scaling resources, violating SLA-compliance.

Phan et al.~\cite{phan2022traffic} proposed a reactive auto-scaling solution for edge deployments for IoT devices which dynamically allocates resources based on incoming traffic. This traffic-aware horizontal pod autoscaler (THPA) operates on top of the underlying Kubernetes architecture. The default Kubernetes horizontal pod autoscaler scales resources in a round-robin manner, not taking into context which nodes are receiving the highest resource requests. THPA alleviates this issue by modelling the resource requests per Kubernetes nodes and intelligently allocating pods to the nodes with higher number of requests. The authors demonstrated that this approach provided a 150\% improvement in response time and throughput. However the algorithm is not SLA compliant due to the delay in scaling resources in a reactive manner.

\subsection{Proactive Autoscaling Strategies}
\label{subsec:proactive}

Lorido et al.~\cite{lorido2014review} showed that compared to reactive algorithms, proactive algorithms achieved better resource allocation once they had been carefully optimized. Machine learning techniques such as auto-regressive integrated moving averages (ARIMA) and long short-term memory (LSTM) have gained popularity in time-series analysis due to their relative ease of building and efficiency compared to other ML models.

Ju et al.~\cite{ju2021proactive} presented a proactive horizontal pod auto-scaling solution for edge computing paradigms. The algorithm, known as Proactive Pod Autoscaler (PPA) was designed to predict resource requests on multiple user-defined metrics, such as CPU request and I/O traffic requests. The algorithm does not use any specific machine learning model for the time-series analysis, instead the model is to be inputted by the user. This model agnostic architecture allows for a very high level of customization. However, such customizability leads to a complex deployment and hyper-parameter tuning process. This, along with a lack of initial training data causes erroneous predictions before the model corrects itself.

Meng et al.~\cite{meng2016crupa} created a proactive auto-scaling algorithm for forecasting the Kubernetes CPU usage of containers using a time-series prediction with ARIMA. The authors demonstrated that such an architecture reduced the forecast errors to 6.5\%, as compared to the baseline of 17\%. However the cost of training this model was prohibitively high, making it unsuitable for edge deployments.

Imdoukh et al.~\cite{imdoukh2020machine} proposed a proactive auto-scaling solution using an LSTM model, designed for edge computing architectures. The algorithm uses an LSTM neural network to predict future network traffic workload to determine the resources to assign to edge nodes ahead of time (cold-start). The authors demonstrated that their algorithm was as accurate as existing ARIMA-based proactive solutions, but significantly reduced the prediction time, as well as computed the minimum resource allocation required to handle future workload. However, this algorithm also suffers from the problems related to a lack of initial training data.

\subsection{Hybrid Autoscaling Strategies}
\label{subsec:hybrid}

All the reactive and proactive approaches have their benefits and drawbacks. Thus, hybrid solutions which merge multiple auto-scaling methods were proposed~\cite{qu2018auto}. While hybrid algorithms for cloud-based deployments exist, integrating them into edge architectures pose several challenges due to the lower data storage and computational capacity of the edge layer. Furthermore, extracting the proactive time-series analysis to the cloud layer poses further challenges due to the inherent latency present between the two layers. Table~\ref{tab:hybrid-autoscalers} shows an overview of the existing proposals compared with our solution.

\begin{table}[t]
    \caption{Summary of hybrid auto-scaling solutions}\label{tab:hybrid-autoscalers}
    \centering
    \small
    \begin{tabular}{@{}lcccccl@{}}
         \toprule
         \multirow{2}{*}{\textbf{Features}}&\multicolumn{5}{c}{\textbf{Hybrid Algorithms}}&\multirow{2}{*}{\textbf{Proposed}}\\
         \cmidrule{2-6}
         &\cite{xu2007use}&\cite{lama2009efficient}&\cite{ramperez2021flas}&\cite{biswas2017hybrid}&\cite{singh2021rhas}&\\
         \midrule
         Simple deployment & \cmark & \cmark & \cmark & \cmark & \cmark & \cmark\\
         Simple parameter tuning & \cmark & \cmark & \cmark & \xmark & \xmark & \cmark\\
         Custom metrics & \cmark & \xmark & \xmark & \cmark & \cmark & \cmark\\
         Light-weight deployment & \cmark & \xmark & \cmark & \xmark & \xmark & \cmark\\
         Edge architecture compliant & \xmark & \xmark & \xmark & \xmark & \xmark & \cmark\\
         SLA-compliant & \xmark & \xmark & \cmark & \cmark & \cmark & \cmark\\
         Minimizes deployment cost & \xmark & \xmark & \xmark & \xmark & \cmark & \cmark\\
         \bottomrule
    \end{tabular}
\end{table}

In 2007, one of the first hybrid algorithms for a distributed deployment was proposed by Jing et al.~\cite{xu2007use}. This algorithm combined rule-based fuzzy inference with machine learning forecasting for dynamic resource allocation. Based on this work, Lama and Zhou~\cite{lama2009efficient} proposed a resource provisioning algorithm for multi-cluster setups using a hybrid autoscaler comprising of a combination of fixed fuzzy rule-based logic and a self adaptive algorithm which dynamically tuned the scaling factor.

Ramp{\'e}rez et al.~\cite{ramperez2021flas} proposed a hybrid approach called Forecasted Load Auto-scaling (FLAS), which combines a predictive model for forecasting time-series resources, while the reactive model estimates other high-level metrics. The linear regression forecaster was however too simplistic to predict complex time-series. Biswas et al.~\cite{biswas2017hybrid} presented a hybrid algorithm designed for cloud computing deployments with service level agreements using an SVM model. Such an SVM-based model is expensive to train however, making it infeasible to deploy on edge deployments.

Singh et al.~\cite{singh2021rhas} proposed another cloud computing based autoscaler with SLA-constraints. The robust hybrid autoscaler (RHAS) was designed particularly for web applications using a modification of the ARIMA machine learning model (TASM). The technique was demonstrated to reduce cloud deployment cost and SLA violations. However, the TASM forecaster was too complex and resource intensive to be deployed on scarce resource paradigms such as the edge layer. This resource intensiveness increased the training times drastically, making it infeasible for conforming to SLA constraints on edge architectures.

The RHAS algorithm by Singh et al.~\cite{singh2021rhas} provided the best approach template for creating a hybrid autoscaler. Therefore, in implementing the autoscaler presented in this paper, we extended the generalized architecture of RHAS to streamline both the reactive and proactive autoscalers, while choosing a more efficient and cost-effective forecasting model to make it SLA-compliant on edge architectures, and eliminating the costly and time-consuming hyper-parameter tuning process.

\section{Problem Formulation}
\label{sec:problem-formulation}

\begin{figure}[htb]
    \centering
    \caption{Autoscaling problem overview}
    \includegraphics[width=1.0\linewidth]{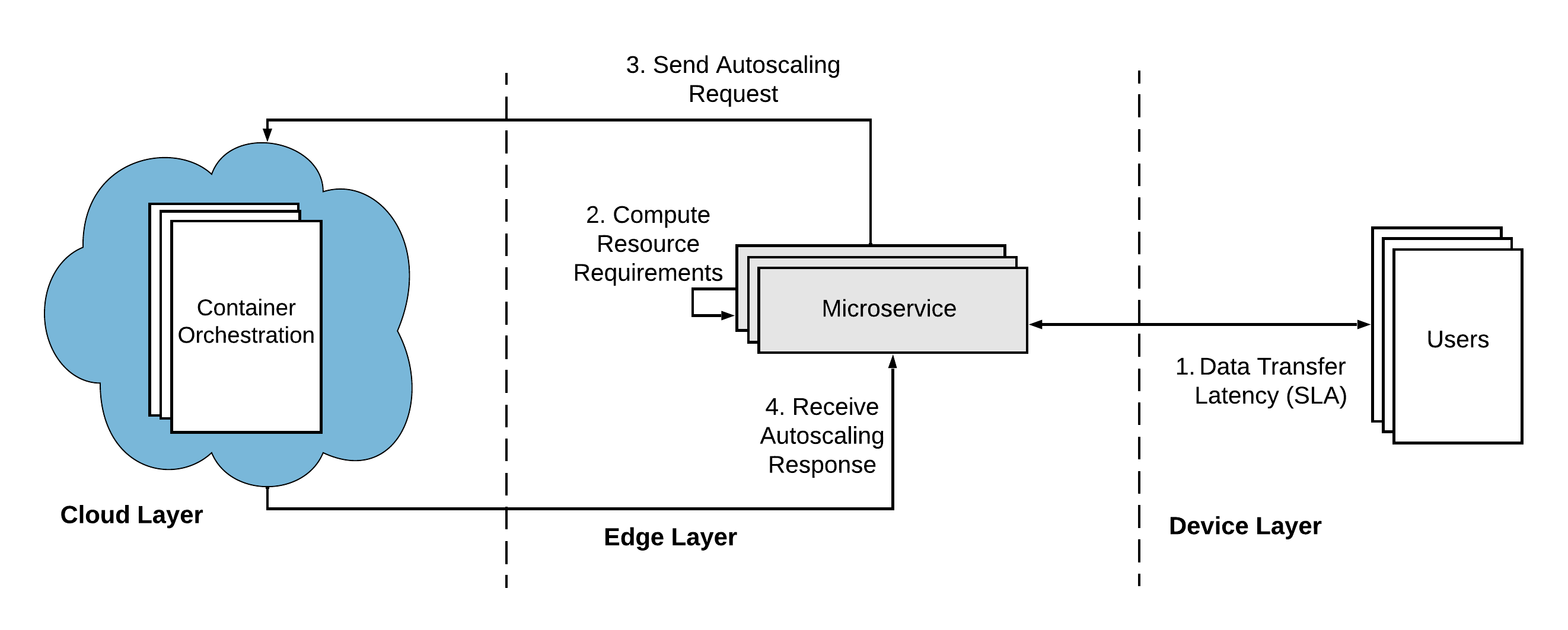}
    \label{fig:autoscaling-problem-overview}
\end{figure}

Figure~\ref{fig:autoscaling-problem-overview} shows the edge architecture layout. The cloud layer is similar to cloud-computing paradigms, wherein it manages the entire network architecture and stores large scale data. The edge layer consists of smaller scale user data storage and communication with user devices. Finally, the device layer consists of all the user devices that will interact with the edge architecture.

The cloud layer has the most amount of resources allocated to it, which it requires when managing the entire network, computing the intensive processing of large-scale data, and coordinating the resource allocation of the edge layer. Only system-critical applications such as the controller orchestration control plane are deployed on this layer. The edge layer has far fewer resources than the cloud layer, but its proximity to the users results in lower network latency, making it ideal for resource scaling. For this reason, the edge layer consists of the orchestration tool's worker nodes and the microservice which receives and serves user data. These worker nodes allocate resources to the microservice deployments dynamically according to user requirements through the process known as auto-scaling.

\subsection{SLA Constraint Definition}
\label{subsec:sla-definition}

Cloud deployments provide several Quality of Service (QoS) metrics when considering SLA negotiations~\cite{serrano2016sla}. These can be broadly classified into performance metrics (response time, throughput), availability metrics (abandon rate, use rate), reliability metrics (mean failure time, mean recovery time), and cost metrics (financial and energy costs).

For this research, we utilize the performance metric \textit{response time} of the user requests as the SLA constraint metric. This metric was chosen due to it being affected the most by intelligently auto-scaling cloud services. By using this metric, the cloud deployment guaranteed that all requests would be served under a certain threshold.

For real-time applications, the auto-scaling should adhere to the SLA metric as much as possible, and try to minimize the number of violations. An SLA constraint $\mathcal{S}_{c}(t)$ is defined as a metric value not exceeding above a threshold $\Delta$ agreed by both the cloud provider and the customer:
\begin{equation}
    \mathcal{S}_{c}(t) > \Delta
    \label{eqn:sla-threshold}
\end{equation}

Following Hussain et al.~\cite{hussain2016sla}, we consider multiple SLA thresholds for different use cases: \textit{Flexible} (highest threshold, typical IoT applications), \textit{Moderate} (trade-off for real-time capabilities), and \textit{Strict} (lowest threshold for time-critical applications such as medical surgeries).

\subsection{Cold-Start Problem}
\label{subsec:cold-start}

The auto-scaling will use a resource metric to scale resources up or down. A problem arises in the time it takes to scale these resources. This time to increase the number of resource replicas $\mathcal{R}$, which we define as the cold start time $\mathcal{C}(t)$:
\begin{equation}
    \mathcal{C}(t) = \mathcal{R}_{deploy}(t) + \mathcal{R}_{register}(t)
\end{equation}
where the cold start time is the summation of the time taken for the replica to be deployed on the data plane and registered with the control plane. The replica image download time is typically a one-time delay due to optimizations done on modern container orchestration software and can be ignored for SLA latency calculations.

When computing the SLA constraint value for a latency metric, the SLA latency can be written as the sum of the cold-start time and the round-trip time taken for the request:
\begin{equation}
    \mathcal{S}_{c}(t) = \mathcal{C}(t) + \mathcal{K}(t) + \frac{\mathcal{U}(t)}{\sum_{i} p_{i}}
    \label{eqn:sla-cold-start}
\end{equation}
where $\mathcal{K}(t)$ is the constant network latency, $\mathcal{U}(t)$ is the maximum latency of a unitary resource deployment, and $\sum_i p_i$ is the total pod count in deployment $\mathcal{D}$. The number of SLA violations $\mathcal{V} \propto \mathcal{C}(t)$ due to the correlation between cold-start delay and the lack of available resources~\cite{patel2021systematic}.

\subsection{Optimization Problem}
\label{subsec:optimization}

From Equation~\ref{eqn:sla-cold-start}, it is clear that $\lim_{\sum_{i} p_{i} \to \infty} \frac{\mathcal{U}(t)}{\sum_{i} p_{i}} = 0$. This incentivizes ignoring intelligent auto-scaling and simply allocating maximum pods. However, most cloud providers allocate a cost for each resource assignment:
\begin{equation}
    cost = \alpha \times \sum_{i} p_{i}
    \label{eqn:cost}
\end{equation}
where $\alpha$ is the unitary resource cost. The auto-scaling optimization problem $\mathcal{P}$ can be formed:
\begin{equation}
    \mathcal{P} = x \times \mathcal{S}_{c}(t) + y \times cost
    \label{eqn:optimization-problem}
\end{equation}
The objective is to minimize both latency and cost. For this research, we configure $x = y = 0.5$, implying both are equally important. Maximizing resources in $\mathcal{D}$ while limiting cost below a threshold to reduce SLA constraint metric is akin to the famous Knapsack Problem~\cite{kellerer2004introduction}. This problem is proven to be NP-Hard, and as such no known algorithm can determine the best value in polynomial time. However, an approximation close to this best value can be computed in polynomial time through reactive rule-based or proactive machine-learning techniques.

\section{Proposed Hybrid Autoscaler}
\label{sec:hybrid-design}

\subsection{Architecture Overview}
\label{subsec:architecture}

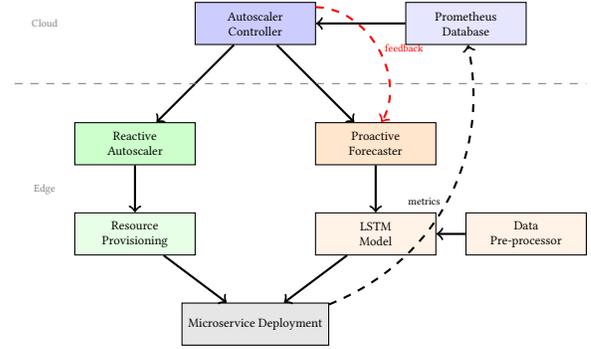
\begin{figure}[t]
    \centering
    \begin{tikzpicture}[
        box/.style={rectangle, draw, minimum width=2cm, minimum height=0.7cm, align=center, font=\scriptsize},
        arrow/.style={->, thick},
        scale=0.8, transform shape
    ]
    \node[box, fill=blue!20] (controller) at (0, 3) {Autoscaler\\Controller};
    \node[box, fill=blue!10] (prometheus) at (3.5, 3) {Prometheus\\Database};
    
    \node[box, fill=green!20] (reactive) at (-2, 1) {Reactive\\Autoscaler};
    \node[box, fill=green!10] (reactive_prov) at (-2, -0.5) {Resource\\Provisioning};
    
    \node[box, fill=orange!20] (proactive) at (2, 1) {Proactive\\Forecaster};
    \node[box, fill=orange!10] (lstm) at (2, -0.5) {LSTM\\Model};
    \node[box, fill=orange!10] (preprocess) at (4.5, -0.5) {Data\\Pre-processor};
    
    \node[box, fill=gray!20] (microservice) at (0, -2) {Microservice Deployment};
    
    \draw[arrow] (controller) -- (reactive);
    \draw[arrow] (controller) -- (proactive);
    \draw[arrow] (prometheus) -- (controller);
    \draw[arrow] (reactive) -- (reactive_prov);
    \draw[arrow] (proactive) -- (lstm);
    \draw[arrow] (preprocess) -- (lstm);
    \draw[arrow] (reactive_prov) -- (microservice);
    \draw[arrow] (lstm) -- (microservice);
    \draw[arrow, dashed] (microservice) to[bend right=40] node[left, font=\tiny] {metrics} (prometheus);
    \draw[arrow, dashed, red] (controller) to[bend left=60] node[right, font=\tiny, red] {feedback} (proactive);
    
    \node[font=\tiny, gray] at (-3.5, 3) {Cloud};
    \node[font=\tiny, gray] at (-3.5, 0.25) {Edge};
    \draw[dashed, gray] (-4, 2) -- (5.5, 2);
    \end{tikzpicture}
    \caption{Proposed hybrid architecture overview}
    \label{fig:hybrid-arch}
\end{figure}

An overview of the hybrid autoscaler architecture is shown in Figure~\ref{fig:hybrid-arch}. The overall architecture is formulated using a hierarchical model. The edge node consists of three main sections. The first is the reactive auto-scaling subsystem, which has the resource provisioning module, and the configuration which dictates the cooldown logic for scaling up and down. As Zhang et al.~\cite{zhang2019quantifying} demonstrated, the microservice system stability is directly related to the careful selection of cool-down parameters.

The second subsystem is the proactive autoscaler. From a high-level perspective, there are three main components. The resource provisioning module is similar to that of the reactive autoscaler, however, it also consists of a forecaster using a deep-learning-based machine learning model, and a data pre-processing algorithm. The data pre-processing algorithm removes any noise present in the time series data, and smoothens the data curves, making it easier for the forecaster to make predictions in a low-cost manner.

Finally, the auto-scaling controller determines which auto-scaling logic will be applied to the replicas, and also keeps track of any SLA violations. It hosts the time-series metric data and has a feedback loop with the proactive autoscaler. If it detects SLA violations during a configured time window, it automatically adjusts the hyper-parameters of the proactive forecaster. Such a heuristic method eliminates the complex hyper-parameter tuning process seen in most proactive models.

\subsection{Scheduler Algorithm}
\label{subsec:scheduler}

At a high level, a container orchestration's default horizontal pod autoscaler operates on the ratio between the current and desired metric values:
\begin{equation}
    replicas_{desired} = \lceil replicas_{current} \times \frac{metric_{current}}{metric_{desired}}\rceil
    \label{eqn:replica-desired}
\end{equation}

The autoscaler controller consists of a scheduling logic module which handles when to switch between proactive and reactive auto-scaling. Algorithm~\ref{alg:scheduler} explains this logic. The autoscaler computes two replica values, one for the proactive forecaster which determines the replicas after $\mathcal{T}$ seconds, and one for the reactive forecaster for current requirements. If the forecasted requirement is higher than current, the scheduler outputs the forecaster replica count as desired replicas. Otherwise, the reactive replica count is used.

\begin{algorithm}[t]
    \caption{Hybrid Scheduler Algorithm}
    \label{alg:scheduler}
    \small
    \textbf{Input}: $replicas_{current}, metric_{current}, metric_{desired}, metric_{forecast}$\\
    \textbf{Output}: $replicas_{desired}$
    \begin{algorithmic}[1]
        \State $replicas_{forecast} \gets \lceil replicas_{current} \times \frac{metric_{forecast}}{metric_{desired}}\rceil$
        \State $replicas_{reactive} \gets \lceil replicas_{current} \times \frac{metric_{current}}{metric_{desired}}\rceil$
        \If{$replicas_{forecast} > replicas_{reactive}$}
            \State $replicas_{desired} \gets replicas_{forecast}$
        \Else
            \State $replicas_{desired} \gets replicas_{reactive}$
        \EndIf
        \State \Return $replicas_{desired}$
    \end{algorithmic}
\end{algorithm}

\subsection{Reactive Resource Provisioning}
\label{subsec:reactive-design}

The reactive autoscaler subsystem is responsible for determining whether auto-scaling should proceed based on given configuration. The reactive algorithm's resource provisioning is built on top of the default horizontal pod autoscaler deployed by Kubernetes. The autoscaler is modified such that it has cooldown parameters set to a moderate value to ensure adaptability to SLA-constrained scenarios while maintaining system stability.

There are three important parameters key to controlling horizontal pod scaling: ``tolerance'', ``scale up cooldown'', and ``scale down cooldown''. The tolerance informs the autoscaler when to skip calculating new replicas:
\begin{equation}
    tolerance = \left| \frac{metric_{desired} - metric_{current}}{metric_{desired}} \right|
    \label{eqn:tolerance}
\end{equation}
By default, if tolerance is below 0.1, autoscaling is skipped. The scale-up and scale-down cooldowns control how quickly auto-scaling occurs. For the proposed autoscaler, both cooldowns are modified to 15 seconds to ensure moderate cooldown values for best system stability and SLA compliance.

\subsection{Data Pre-Processor}
\label{subsec:preprocess}

To speed up the forecast process and reduce resource requirements, the time-series data is pre-processed to smoothen it. This makes it easier for the deep learning model to extract patterns, reduces training and validation loss, and reduces the length of the training window data sequence the LSTM requires. The smoothing is achieved using the Savitzky-Golay filter~\cite{savitzky1964smoothing}, which takes $N$ points in a given time-series, with a filter width $w$, and calculates a polynomial average of order $o$~\cite{schafer2011savitzky}. The resulting data has considerably less deviations between consecutive points and is devoid of noise.

\subsection{Proactive Forecaster}
\label{subsec:proactive-design}

Several time series forecaster algorithms exist, with LSTM and ARIMA being prominent ones. Siami-Namini et al.~\cite{siami2018comparison} demonstrated that LSTM implementations outperformed ARIMA, reducing error rates by over 80\%. Furthermore, the number of training ``epochs'' did not need to be set to a high value; setting significantly higher values degraded performance due to over-fitting. LSTM works well due to ``rolling updates'' on the model---weights are only set once when deployed, then always updated on every training call.

Algorithm~\ref{alg:proactive-forecast} shows the forecaster implementation. The autoscaler controller implements a control loop every $\mathcal{P}$ seconds requesting the latest prediction. The forecaster pre-processes data to remove noise, performs training with configured hyper-parameters, computes validation loss, and accepts this model if it has lower validation loss than previous iterations. Finally, the model predicts future metrics and returns them to the controller.

\begin{algorithm}[t]
    \caption{Proactive Forecaster Algorithm}
    \label{alg:proactive-forecast}
    \small
    \textbf{Input}: $lookback \geq 0, epochs \leq 100, learning\_rate \leq 1$\\
    \textbf{Output}: $metric_{forecast}$
    \begin{algorithmic}[1]
        \State $lstm\_model \gets lstm.initialize()$
        \State $time\_series \gets get\_latest\_data()$
        \State $lstm\_input \gets get\_input(time\_series, lookback)$
        \State $lstm\_input \gets preprocess\_data(lstm\_input)$
        \State $new\_model \gets train(lstm\_input, epochs, learning\_rate)$
        \If{$validation\_loss(new\_model) < validation\_loss(lstm\_model)$}
            \State $lstm\_model \gets new\_model$
        \EndIf
        \State $metric_{forecast} \gets lstm\_model.predict(lstm\_input)$
        \State \Return $metric_{forecast}$
    \end{algorithmic}
\end{algorithm}

The proactive forecaster is a deep-learning model configured with multi-step forecast output. The model consists of three LSTM layers, alternated with two dropout layers (to prevent over-fitting), and a final densely connected neural-network generating the forecaster output. The output is 540 data points (approximately 24 hours of workload), so the forecaster only needs to run once daily, vastly reducing total training time.

\begin{table}[t]
    \caption{Proactive forecaster layer configuration}\label{tab:lstm-layers}
    \centering
    \small
    \begin{tabular}{lcc}
        \toprule
        \textbf{Layer Details} & \textbf{Output Shape} & \textbf{Parameters}\\
        \midrule
        LSTM$_1$ & (10, 50) & 10,400\\
        Dropout$_1$ & (10, 50) & 0\\
        LSTM$_2$ & (10, 50) & 20,200\\
        Dropout$_2$ & (10, 50) & 0\\
        LSTM$_3$ & (50) & 20,200\\
        Dense$_1$ & (540) & 27,540\\
        \midrule
        \textbf{Total} & & \textbf{78,340}\\
        \bottomrule
    \end{tabular}
\end{table}

Default hyper-parameters are: learning rate = 0.005, epochs = 75, batch size = 100, with Adam optimizer~\cite{diederik2014adam}. An ``early-stop'' function halts training if loss does not decrease for 10 consecutive epochs. The model training, validation, and error comparison takes approximately 3 minutes, after which the model predicts the subsequent day's forecast in under 10 seconds.

\subsection{SLA-based Heuristic Feedback}
\label{subsec:feedback}

The autoscaler controller constantly checks for SLA violations using a control loop. Typically, SLA checks are done for a sufficiently lengthy period such as one day. If an SLA violation is found, it is concluded that the application was unable to autoscale quickly enough to avoid the cold start problem. This could be due to insufficient training data or conservative hyper-parameter selections.

To temporarily boost learning, the controller decreases the learning rate (to increase probability of escaping local minima), increases batch size (to reduce under-fitting), and increases epochs (to reduce loss). All parameters have thresholds to prevent over-fitting or infeasibly lengthy training times. Algorithm~\ref{alg:feedback} shows this implementation.

\begin{algorithm}[t]
    \caption{SLA-based Heuristic Feedback}
    \label{alg:feedback}
    \small
    \textbf{Input}: $\mathcal{V}, learning\_rate, batch\_size, epochs$\\
    \textbf{Output}: $hyperparameters_{modified}$
    \begin{algorithmic}[1]
        \State $initial\_rate \gets learning\_rate$
        \State $initial\_batch \gets batch\_size$
        \State $initial\_epochs \gets epochs$
        \If{$\mathcal{V} > 0$}
            \State $batch\_size \gets \min(batch\_size + 10, 200)$
            \State $learning\_rate \gets \max(learning\_rate - 0.0005, 0.002)$
            \State $epochs \gets \min(epochs + 5, 100)$
        \Else
            \State Reset to initial values
        \EndIf
        \State \Return $(learning\_rate, batch\_size, epochs)$
    \end{algorithmic}
\end{algorithm}

If the feedback control loop discovers no SLA-violations during a time-period, it concludes that the LSTM has sufficiently learned the primary characteristics. The ``rolling-updates'' feature of LSTM allows safely resetting hyper-parameters while preserving learning and weights of previous training rounds.

\subsection{Complexity Analysis}
\label{subsec:complexity}

Assuming the hybrid autoscaler $\mathcal{H}$ takes time-series data of length $\mathcal{N}$, stored in an array data structure, and LSTM weights as a two-dimensional matrix of size $A \times B$:

\textbf{Space Complexity}: The time-series array has complexity $O(N)$ and weights are $O(N^2)$. Thus:
\begin{equation}
    Complexity_{space}(\mathcal{H}) = O(N^2)
\end{equation}

\textbf{Time Complexity}: The reactive autoscaler and controller only compute tolerance values---constant operations: $O(1)$. For the proactive autoscaler, the LSTM internally computes matrix multiplications. For dimensions $m$ and $n$:
\begin{equation}
    Complexity_{time}(proactive) = O(m \times (m + n + 1)) = O(N^2)
\end{equation}

For $\tau$ training epochs (a constant), the final complexity remains $O(N^2)$. Combining all components:
\begin{equation}
    Complexity_{time}(\mathcal{H}) = O(1) + O(1) + O(N^2) = O(N^2)
\end{equation}

The hybrid algorithm performs in polynomial time complexity, providing an approximation for the NP-Hard optimization problem.

\section{Experimental Setup}
\label{sec:experimental-setup}

\subsection{Cluster Configuration}
\label{subsec:cluster}

For the underlying virtual machine (VM) setup, servers in a private cloud were leveraged. The setup consisted of 6 VMs, using a total of 24 CPU cores and 80GB of memory. These servers were separated into a cloud and an edge layer. The servers on the cloud layer have substantially higher CPU cores and memory compared to the edge layer, to simulate resource scarcity in the edge layer. The cloud layer also contained a 200GB persistent storage volume for Prometheus data, while the edge layer stored time series in RAM. A simulated latency was added between inter-layer communication to mimic perceived distance between edge nodes and data centers.

\begin{table}[t]
    \caption{Cluster architectural layout}\label{tab:cluster-hw}
    \centering
    \small
    \begin{tabular}{lccc}
        \toprule
        \textbf{Node} & \textbf{Layer} & \textbf{CPU} & \textbf{Memory}\\
        \midrule
        Control-Plane-K8s & Cloud & 8 cores & 32GB\\
        Control-Plane-DB  & Cloud & 8 cores & 32GB\\
        Data-Plane-1      & Edge  & 2 cores & 4GB\\
        Data-Plane-2      & Edge  & 2 cores & 4GB\\
        Data-Plane-3      & Edge  & 2 cores & 4GB\\
        Data-Plane-4      & Edge  & 2 cores & 4GB\\
        \bottomrule
    \end{tabular}
\end{table}

Each server uses Ubuntu 22.04. Kubernetes v1.28.2 is used as the container orchestration technology. CRI-O was installed as the container runtime, and Flannel for inter-pod communication. A bare-metal Kubernetes implementation was used for maximum flexibility, with the control plane on the cloud layer and data plane on the edge layer.

\subsection{Benchmark Application}
\label{subsec:benchmark}

DeathStarBench~\cite{gan2019open}, a social network microservice implementation, was deployed for conducting benchmarks on edge architectures with SLA constraints. The application mimics a typical large-scale social network supporting common actions: registering and login, creating user posts, reading timelines, receiving follower recommendations, following/unfollowing users, and searching.

The end-to-end service uses HTTP requests processed by NGINX load balancer, which communicates with microservices in the logic layer for composing and displaying user and timeline posts. The logic layer handles posts containing text, links, and media. Results are stored using memcached for caching and MongoDB for persistent storage.

Based on the wrk2 benchmark, two APIs were identified for testing. One was a GET call to user's home timeline (\textit{home-timeline-service}), and the other was a POST request for creating posts (\textit{compose-post-service}). These were identified as bottlenecks through Jaeger tracing, making them prime targets for auto-scaling.

\subsection{Workload Generation}
\label{subsec:workload}

The social media deployment comes with an HTTP workload generator, wrk2, which creates realistic simulation of typical daily workload. A typical IoT application in the edge has a semi-predictable workload pattern. Tadakamalla and Menasc{\'e}~\cite{tadakamalla2019characterization} demonstrated through a survey that IoT application workloads can be well approximated using a lognormal distribution, and daily routines of users greatly affect workload patterns.

The workload assumes peaks in morning and evening, moderate usage during afternoon, and lowest at night. The workload simulator was modified to introduce randomness to mimic realistic weekly workloads, varying on occasions such as weekends and holidays. A total of approximately 2,550,000 requests were sent over five days per experiment.

\subsection{Baseline Algorithms}
\label{subsec:baselines}

Three baseline algorithms were chosen for comparison, all auto-scaling at the same CPU threshold:

\begin{enumerate}
    \item \textbf{Default Kubernetes HPA}: No modifications---scale-up cooldown is 0 seconds, scale-down is 300 seconds. No knowledge of workload distribution or SLA violations on edge nodes.
    
    \item \textbf{Traffic Aware HPA (THPA)}~\cite{phan2022traffic}: Computes ratio of workloads on different edge nodes with deployment pods, scaling resources in commensurate proportion.
    
    \item \textbf{Proactive Pod Autoscaler (PPA)}~\cite{ju2021proactive}: Uses an LSTM model similar to our hybrid autoscaler but without pre-processing, dealing with more complex time-series data, requiring deeper architecture. The LSTM continuously loops through time-series data and saves forecast results. An update loop updates the model using latest forecasts. Hyper-parameters are carefully tuned but no SLA feedback is provided.
\end{enumerate}

\subsection{SLA Thresholds}
\label{subsec:sla-thresholds}

According to Nilsson and Yngwe~\cite{nilsson2022api}, user experience is negatively affected by higher API latency. Their research found three latency brackets: $\leq$100ms (instantaneous), $\leq$1 second (slight delay), and $>$10 seconds (user loses focus). Based on this, we defined three SLA categories (Table~\ref{tab:sla-values}).

\begin{table}[t]
    \caption{Experimental SLA constraints}\label{tab:sla-values}
    \centering
    \small
    \begin{tabular}{lcc}
        \toprule
        \textbf{SLA Type} & \textbf{GET latency (ms)} & \textbf{POST latency (ms)}\\
        \midrule
        Flexible    & 150   & 1000\\
        Moderate    & 125   & 900\\
        Strict      & 100   & 800\\
        \bottomrule
    \end{tabular}
\end{table}

\section{Performance Evaluation}
\label{sec:evaluation}

Two independent experiments were conducted to validate the hybrid autoscaler performance. The social media application was first tested using GET requests to autoscale \textit{home-timeline-service}. Then, a more demanding workload was applied using POST requests for \textit{compose-post-service}. Both experiments used the workload generation algorithm over five days.

\subsection{Request Latency Analysis}
\label{subsec:latency-eval}

\subsubsection{Default Kubernetes Autoscaler Baseline}

\begin{figure}[t]
    \centering
    \begin{tikzpicture}
        \begin{axis}[
            height=4cm,
            width=\columnwidth,
            xmin=0, xmax=2460,
            ymin=0, ymax=500,
            xlabel={Time (data points)},
            ylabel={Latency (ms)},
            legend style={at={(0.5,1)}, anchor=north, legend columns=2, font=\tiny},
            grid=major,
            grid style={dashed, gray!30},
        ]
            \addplot[RoyalBlue, thick, smooth, mark=none] table [x=xtick, y=value, col sep=comma] {Data/Home-Timeline-Default-Latency-CPU.csv};
            \addplot[OliveGreen, thick, smooth, mark=none] table [x=xtick, y=value, col sep=comma] {Data/Home-Timeline-Default-Latency-AVG.csv};
            \draw[red, thick, dashed] (axis cs:0,150) -- (axis cs:2460,150);
            \legend{Latency, Cumulative Avg, SLA Threshold}
        \end{axis}
    \end{tikzpicture}
    \caption{Default Kubernetes autoscaler latency for GET requests}
    \label{fig:exp1-default}
\end{figure}
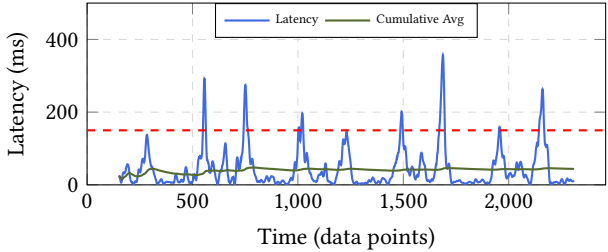

Figure~\ref{fig:exp1-default} shows the default Kubernetes HPA results for GET requests. The autoscaler was merely a primitive reactive implementation with no knowledge of which edge nodes experienced heavy traffic. Thus, it blindly assigned pods in a round-robin manner. Additionally, the autoscaler required significant time to register new pods, falling victim to the cold start problem. This results in significant latency spikes before resources are adjusted. The latency exceeded 300ms at some points, significantly large enough to degrade user experience. By the fifth day, cumulative average latency was nearly 50ms.

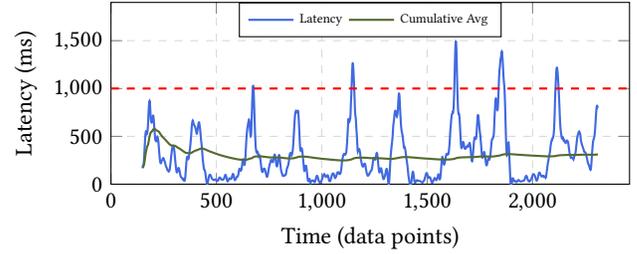
\begin{figure}[t]
    \centering
    \begin{tikzpicture}
        \begin{axis}[
            height=4cm,
            width=\columnwidth,
            xmin=0, xmax=2460,
            ymin=0, ymax=1900,
            xlabel={Time (data points)},
            ylabel={Latency (ms)},
            legend style={at={(0.5,1)}, anchor=north, legend columns=2, font=\tiny},
            grid=major,
            grid style={dashed, gray!30},
        ]
            \addplot[RoyalBlue, thick, smooth, mark=none] table [x=xtick, y=value, col sep=comma] {Data/Compose-Post-Default-Latency-CPU.csv};
            \addplot[OliveGreen, thick, smooth, mark=none] table [x=xtick, y=value, col sep=comma] {Data/Compose-Post-Default-Latency-AVG.csv};
            \draw[red, thick, dashed] (axis cs:0,1000) -- (axis cs:2460,1000);
            \legend{Latency, Cumulative Avg, SLA Threshold}
        \end{axis}
    \end{tikzpicture}
    \caption{Default Kubernetes autoscaler latency for POST requests}
    \label{fig:exp2-default}
\end{figure}

For POST requests (Figure~\ref{fig:exp2-default}), the shortcomings were exposed even more by increased demands. During daily workload spikes, the autoscaler regularly breached 1000ms, with values peaking at almost 1450ms---more than 45\% above threshold. Furthermore, the average latency hovered around 400ms throughout. Investigation revealed three issues: (1) cold start problem adding constant latency, (2) avalanche effect from resources not being available timely, causing connections to be dropped and 60-second timeouts, and (3) uneven distribution of requests due to round-robin scheduling.

\subsubsection{Reactive THPA Autoscaler Baseline}

\begin{figure}[t]
    \centering
    \begin{tikzpicture}
        \begin{axis}[
            height=4cm,
            width=\columnwidth,
            xmin=0, xmax=2460,
            ymin=0, ymax=250,
            xlabel={Time (data points)},
            ylabel={Latency (ms)},
            legend style={at={(0.5,1)}, anchor=north, legend columns=2, font=\tiny},
            grid=major,
            grid style={dashed, gray!30},
        ]
            \addplot[RoyalBlue, thick, smooth, mark=none] table [x=xtick, y=value, col sep=comma] {Data/Home-Timeline-Reactive-Latency-CPU.csv};
            \addplot[OliveGreen, thick, smooth, mark=none] table [x=xtick, y=value, col sep=comma] {Data/Home-Timeline-Reactive-Latency-AVG.csv};
            \draw[red, thick, dashed] (axis cs:0,150) -- (axis cs:2460,150);
            \legend{Latency, Cumulative Avg, SLA Threshold}
        \end{axis}
    \end{tikzpicture}
    \caption{THPA reactive autoscaler latency for GET requests}
    \label{fig:exp1-reactive}
\end{figure}
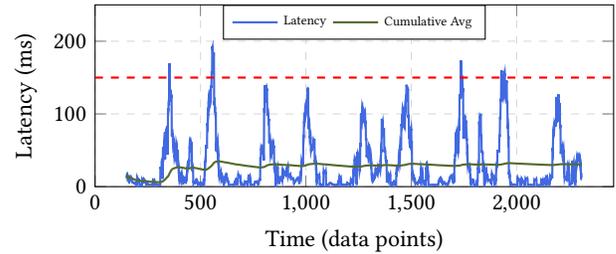

Unlike the default autoscaler, THPA keeps track of which edge nodes receive significant requests and assigns pods accordingly. This resulted in significantly improved latency for GET requests (Figure~\ref{fig:exp1-reactive}). While still suffering from cold start, the more intelligent resource assignment resulted in fewer availability issues. However, the SLA threshold of 150ms was still regularly breached, though breaches never exceeded 200ms. The average latency was 25-30ms.

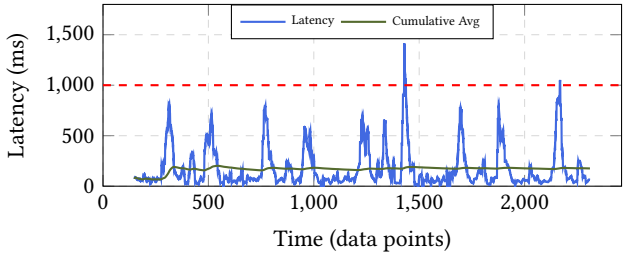
\begin{figure}[t]
    \centering
    \begin{tikzpicture}
        \begin{axis}[
            height=4cm,
            width=\columnwidth,
            xmin=0, xmax=2460,
            ymin=0, ymax=1800,
            xlabel={Time (data points)},
            ylabel={Latency (ms)},
            legend style={at={(0.5,1)}, anchor=north, legend columns=2, font=\tiny},
            grid=major,
            grid style={dashed, gray!30},
        ]
            \addplot[RoyalBlue, thick, smooth, mark=none] table [x=xtick, y=value, col sep=comma] {Data/Compose-Post-Reactive-Latency-CPU.csv};
            \addplot[OliveGreen, thick, smooth, mark=none] table [x=xtick, y=value, col sep=comma] {Data/Compose-Post-Reactive-Latency-AVG.csv};
            \draw[red, thick, dashed] (axis cs:0,1000) -- (axis cs:2460,1000);
            \legend{Latency, Cumulative Avg, SLA Threshold}
        \end{axis}
    \end{tikzpicture}
    \caption{THPA reactive autoscaler latency for POST requests}
    \label{fig:exp2-reactive}
\end{figure}

For POST requests (Figure~\ref{fig:exp2-reactive}), the request-aware architecture eliminated dropped request issues. The avalanche effect was somewhat mitigated. However, cold start still caused spikes above SLA threshold on multiple occasions, with latency nearly hitting 1400ms before correcting. Cumulative average latency was around 200ms---substantially lower than default.

\subsubsection{Proactive PPA Autoscaler Baseline}

The PPA algorithm attempts to predict workload before it is requested, eliminating cold start in ideal conditions. However, experiments showed otherwise. Because the autoscaler was purely proactive, it requires a deep LSTM model with several layers and large training epochs. This deep model took more than 50 minutes to properly train for 24-hour predictions due to edge architecture's lack of resources.

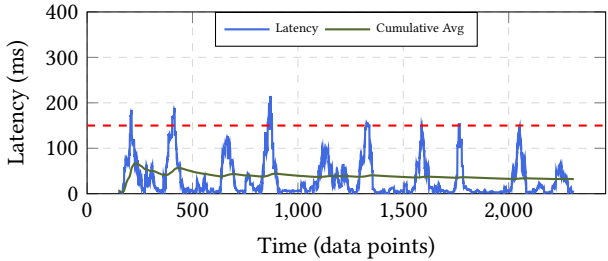
\begin{figure}[t]
    \centering
    \begin{tikzpicture}
        \begin{axis}[
            height=4cm,
            width=\columnwidth,
            xmin=0, xmax=2460,
            ymin=0, ymax=400,
            xlabel={Time (data points)},
            ylabel={Latency (ms)},
            legend style={at={(0.5,1)}, anchor=north, legend columns=2, font=\tiny},
            grid=major,
            grid style={dashed, gray!30},
        ]
            \addplot[RoyalBlue, thick, smooth, mark=none] table [x=xtick, y=value, col sep=comma] {Data/Home-Timeline-Proactive-Latency-CPU.csv};
            \addplot[OliveGreen, thick, smooth, mark=none] table [x=xtick, y=value, col sep=comma] {Data/Home-Timeline-Proactive-Latency-AVG.csv};
            \draw[red, thick, dashed] (axis cs:0,150) -- (axis cs:2460,150);
            \legend{Latency, Cumulative Avg, SLA Threshold}
        \end{axis}
    \end{tikzpicture}
    \caption{PPA proactive autoscaler latency for GET requests}
    \label{fig:exp1-proactive}
\end{figure}

Figure~\ref{fig:exp1-proactive} shows GET request results. Initially, latency continually spiked causing many SLA violations---more than the reactive autoscaler. However, after several days of training, rolling updates stabilized the latency. SLA violations were not as severe as default baseline but comparatively greater than reactive, exceeding 200ms for several minutes daily. Average latency approached 50ms.

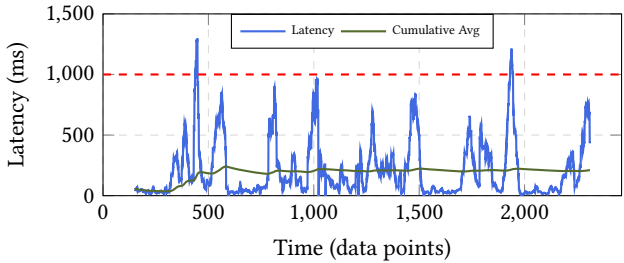
\begin{figure}[t]
    \centering
    \begin{tikzpicture}
        \begin{axis}[
            height=4cm,
            width=\columnwidth,
            xmin=0, xmax=2460,
            ymin=0, ymax=1500,
            xlabel={Time (data points)},
            ylabel={Latency (ms)},
            legend style={at={(0.5,1)}, anchor=north, legend columns=2, font=\tiny},
            grid=major,
            grid style={dashed, gray!30},
        ]
            \addplot[RoyalBlue, thick, smooth, mark=none] table [x=xtick, y=value, col sep=comma] {Data/Compose-Post-Proactive-Latency-CPU.csv};
            \addplot[OliveGreen, thick, smooth, mark=none] table [x=xtick, y=value, col sep=comma] {Data/Compose-Post-Proactive-Latency-AVG.csv};
            \draw[red, thick, dashed] (axis cs:0,1000) -- (axis cs:2460,1000);
            \legend{Latency, Cumulative Avg, SLA Threshold}
        \end{axis}
    \end{tikzpicture}
    \caption{PPA proactive autoscaler latency for POST requests}
    \label{fig:exp2-proactive}
\end{figure}

For POST requests (Figure~\ref{fig:exp2-proactive}), latency initially spiked above 1200ms before stabilizing, with one more spike on the last day. The first spike occurred due to insufficient training data---the complex LSTM without pre-processing made it difficult to correctly predict data curves early. This resolved as more data was added, but a threshold was reached where data was so large that forecasting took significantly longer, causing the final day spike. Average latency was around 200ms.

\subsubsection{Proposed Hybrid Autoscaler}

Finally, with baselines established, the hybrid algorithm was tested. This approach mitigates issues seen in both reactive and proactive approaches. The autoscaler is extremely lightweight and easy to configure since no hyper-parameter tuning is required. The proactive forecaster accurately predicts the beginning of workload spikes, eliminating cold start. The forecaster cannot accurately predict middle and end of daily workloads, but this is not an issue since the reactive algorithm handles these, ensuring SLA compliance.

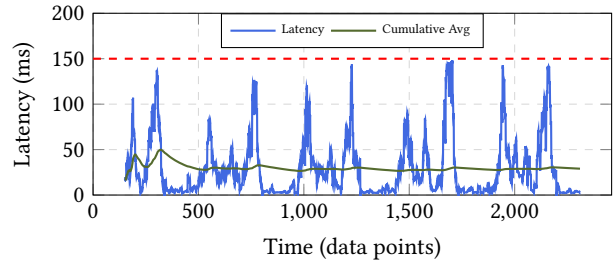
\begin{figure}[t]
    \centering
    \begin{tikzpicture}
        \begin{axis}[
            height=4cm,
            width=\columnwidth,
            xmin=0, xmax=2460,
            ymin=0, ymax=200,
            xlabel={Time (data points)},
            ylabel={Latency (ms)},
            legend style={at={(0.5,1)}, anchor=north, legend columns=2, font=\tiny},
            grid=major,
            grid style={dashed, gray!30},
        ]
            \addplot[RoyalBlue, thick, smooth, mark=none] table [x=xtick, y=value, col sep=comma] {Data/Home-Timeline-Hybrid-Latency-CPU.csv};
            \addplot[OliveGreen, thick, smooth, mark=none] table [x=xtick, y=value, col sep=comma] {Data/Home-Timeline-Hybrid-Latency-AVG.csv};
            \draw[red, thick, dashed] (axis cs:0,150) -- (axis cs:2460,150);
            \legend{Latency, Cumulative Avg, SLA Threshold}
        \end{axis}
    \end{tikzpicture}
    \caption{Hybrid autoscaler latency for GET requests}
    \label{fig:exp1-hybrid}
\end{figure}

Figure~\ref{fig:exp1-hybrid} shows the GET request results. No SLA violations occurred during the five-day workload. The controller did not intervene in LSTM training or modify forecaster hyper-parameters. Average latency was around 30ms, similar to THPA reactive implementation. This performance was a significant improvement over baseline algorithms.

\begin{figure}[t]
    \centering
    \begin{tikzpicture}
        \begin{axis}[
            height=4cm,
            width=\columnwidth,
            xmin=0, xmax=2460,
            ymin=0, ymax=1400,
            xlabel={Time (data points)},
            ylabel={Latency (ms)},
            legend style={at={(0.5,1)}, anchor=north, legend columns=2, font=\tiny},
            grid=major,
            grid style={dashed, gray!30},
        ]
            \addplot[RoyalBlue, thick, smooth, mark=none] table [x=xtick, y=value, col sep=comma] {Data/Compose-Post-Hybrid-Latency-CPU.csv};
            \addplot[OliveGreen, thick, smooth, mark=none] table [x=xtick, y=value, col sep=comma] {Data/Compose-Post-Hybrid-Latency-AVG.csv};
            \draw[red, thick, dashed] (axis cs:0,1000) -- (axis cs:2460,1000);
            \legend{Latency, Cumulative Avg, SLA Threshold}
        \end{axis}
    \end{tikzpicture}
    \caption{Hybrid autoscaler latency for POST requests}
    \label{fig:exp2-hybrid}
\end{figure}
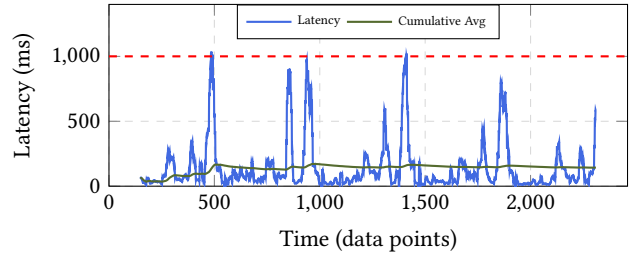

Figure~\ref{fig:exp2-hybrid} shows POST request results. Only one SLA violation occurred on the first day due to lack of training data causing erroneous prediction. However, the reactive subsystem took over and scaled resources accordingly, so the threshold was only breached slightly (peaking at $\sim$1020ms). After this violation, the controller deduced that training needed kick-starting through hyper-parameter tuning. New hyper-parameter values were provided in the next training cycle, and no violations occurred the next day. The controller then reset hyper-parameters to speed up training, and even though latency approached 990ms on the third day, no further violations occurred. Average latency was below 200ms.

The hybrid approach nearly eliminated the cold start problem very early in the experiment. Initial forecasting difficulties were quickly resolved by the controller's corrective instructions. All this was done with no user intervention, making the autoscaler extremely autonomous. The algorithm completed training within a few minutes, allowing quick resource registration. CPU utilization never exceeded 100\%, so no user requests were dropped, allowing full system availability.

\subsection{CPU Workload Distribution}
\label{subsec:cpu-eval}

The distribution of CPU workload across deployment pods is another important metric. The goal is to maximize deployment resources while minimizing costs. Ideally, the distributed workload should hover at $\frac{\mathcal{A}}{2}$ where $\mathcal{A}$ is the autoscaler threshold. When workload approaches $\mathcal{A}$, not enough pods are deployed, causing queued or dropped requests. When workload tends towards 0, too many pods have been assigned, increasing costs with low latency reduction returns.

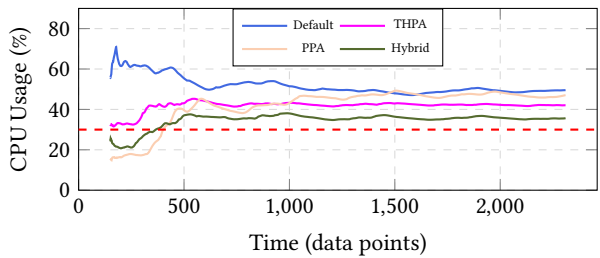
\begin{figure}[t]
    \centering
    \begin{tikzpicture}
        \begin{axis}[
            height=4cm,
            width=\columnwidth,
            xmin=0, xmax=2460,
            ymin=0, ymax=90,
            xlabel={Time (data points)},
            ylabel={CPU Usage (\%)},
            legend style={at={(0.5,1)}, anchor=north, legend columns=2, font=\tiny},
            grid=major,
            grid style={dashed, gray!30},
        ]
            \addplot[RoyalBlue, thick, smooth, mark=none] table [x=xtick, y=value, col sep=comma] {Data/Compose-Post-CPU-Usage-Default.csv};
            \addplot[Magenta, thick, smooth, mark=none] table [x=xtick, y=value, col sep=comma] {Data/Compose-Post-CPU-Usage-Reactive.csv};
            \addplot[Apricot, thick, smooth, mark=none] table [x=xtick, y=value, col sep=comma] {Data/Compose-Post-CPU-Usage-Proactive.csv};
            \addplot[OliveGreen, thick, smooth, mark=none] table [x=xtick, y=value, col sep=comma] {Data/Compose-Post-CPU-Usage-Hybrid.csv};
            \draw[red, thick, dashed] (axis cs:0,30) -- (axis cs:2460,30);
            \legend{Default, THPA, PPA, Hybrid, Ideal}
        \end{axis}
    \end{tikzpicture}
    \caption{CPU workload distribution for POST requests}
    \label{fig:cpu-distribution}
\end{figure}

Figure~\ref{fig:cpu-distribution} shows average CPU utilization for all algorithms. For GET requests (threshold 50\%, ideal 25\%), Default achieved $\sim$35\%, PPA $\sim$33\%, THPA $\sim$30\%, and Hybrid $\sim$26\%. For POST requests (threshold 60\%, ideal 30\%), Default peaked at 70\% before stabilizing at 50\% (with uneven distribution), PPA at $\sim$45\%, THPA at $\sim$42\%, and Hybrid at $\sim$35\%---closest to optimal in both cases.

\subsection{SLA Violation Rates}
\label{subsec:sla-eval}

As demonstrated above, the hybrid autoscaler performed significantly better than baselines with flexible SLA thresholds. For thorough demonstration, all algorithms were tested on moderate and strict thresholds. The workload was run for five days, with auto-scaling performed only for the last two days to ensure best possible results regardless of training data length.

\subsubsection{GET Request SLA Violations}

\begin{figure}[t]
    \centering
    \begin{tikzpicture}
        \begin{axis}[
            ybar,
            height=4.5cm,
            width=\columnwidth,
            bar width=10pt,
            enlarge x limits=0.3,
            xlabel={SLA Category},
            ylabel={Violations (\%)},
            symbolic x coords={Flexible, Moderate, Strict},
            xtick=data,
            ymin=0, ymax=22,
            legend style={at={(0.5,1)}, anchor=north, legend columns=4, font=\tiny},
            nodes near coords,
            nodes near coords style={font=\tiny, rotate=45, anchor=west},
        ]
            \addplot[fill=RoyalBlue] coordinates {(Flexible,5.55) (Moderate,8.74) (Strict,14.8)};
            \addplot[fill=Magenta] coordinates {(Flexible,1.25) (Moderate,7.76) (Strict,12.76)};
            \addplot[fill=Apricot] coordinates {(Flexible,2.36) (Moderate,2.91) (Strict,5.41)};
            \addplot[fill=OliveGreen] coordinates {(Flexible,0.00) (Moderate,0.42) (Strict,1.66)};
            \legend{Default, THPA, PPA, Hybrid}
        \end{axis}
    \end{tikzpicture}
    \caption{SLA violation rates for GET requests}
    \label{fig:sla-get}
\end{figure}
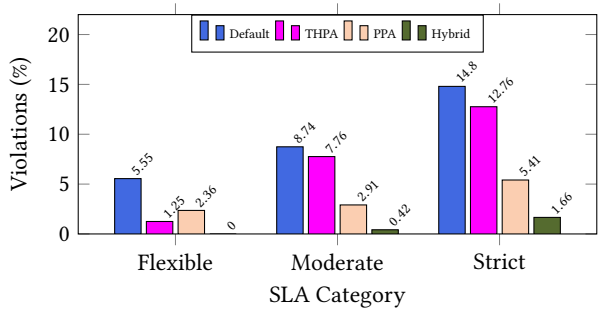

Figure~\ref{fig:sla-get} and Table~\ref{tab:sla-get} show SLA violations for GET requests across all threshold categories.

\begin{table}[t]
    \caption{SLA violation counts for GET requests}\label{tab:sla-get}
    \centering
    \small
    \begin{tabular}{lccc}
        \toprule
        \textbf{Algorithm} & \textbf{Flexible} & \textbf{Moderate} & \textbf{Strict}\\
        \midrule
        Total Requests & 2,550,000 & 1,220,000 & 1,220,000\\
        \midrule
        Default & 141,525 & 106,628 & 180,560\\
        THPA & 31,875 & 94,672 & 155,672\\
        PPA & 60,180 & 35,502 & 66,002\\
        \textbf{Hybrid} & \textbf{0} & \textbf{5,124} & \textbf{20,252}\\
        \bottomrule
    \end{tabular}
\end{table}

For the flexible category, Default performed worst (5.55\%), PPA achieved 2.36\% (increased due to training model complexity), THPA performed well (1.25\%), and Hybrid achieved \textbf{0\%}---qualifying for ``highly available'' SLA deployment.

The moderate threshold proved far more difficult. Default and THPA showed similarly poor results (8.74\% and 7.76\% respectively), demonstrating the importance of mitigating cold start. PPA achieved 2.91\%, while Hybrid still achieved best results with only 0.42\% violations (initial violations quickly counteracted by hyper-parameter tuning).

For the strict threshold, Default and reactive autoscalers performed poorly (14.8\% and 12.76\%). PPA clearly showed cold start importance with only 5.41\%. Hybrid performed substantially better with only 1.66\%.

\subsubsection{POST Request SLA Violations}

\begin{figure}[t]
    \centering
    \begin{tikzpicture}
        \begin{axis}[
            ybar,
            height=4.5cm,
            width=\columnwidth,
            bar width=10pt,
            enlarge x limits=0.3,
            xlabel={SLA Category},
            ylabel={Violations (\%)},
            symbolic x coords={Flexible, Moderate, Strict},
            xtick=data,
            ymin=0, ymax=35,
            legend style={at={(0.5,1)}, anchor=north, legend columns=4, font=\tiny},
            nodes near coords,
            nodes near coords style={font=\tiny, rotate=45, anchor=west},
        ]
            \addplot[fill=RoyalBlue] coordinates {(Flexible,3.61) (Moderate,12.3) (Strict,22.38)};
            \addplot[fill=Magenta] coordinates {(Flexible,0.56) (Moderate,9.58) (Strict,18.8)};
            \addplot[fill=Apricot] coordinates {(Flexible,0.61) (Moderate,7.22) (Strict,9.94)};
            \addplot[fill=OliveGreen] coordinates {(Flexible,0.13) (Moderate,3.34) (Strict,5.41)};
            \legend{Default, THPA, PPA, Hybrid}
        \end{axis}
    \end{tikzpicture}
    \caption{SLA violation rates for POST requests}
    \label{fig:sla-post}
\end{figure}
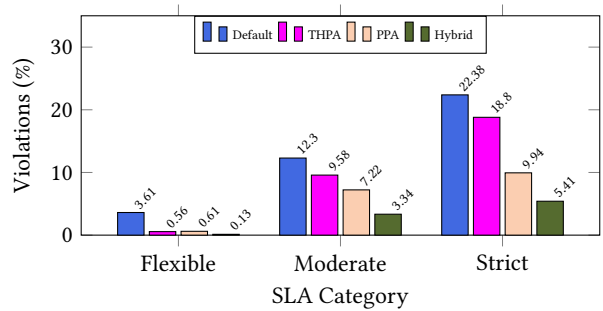

Figure~\ref{fig:sla-post} and Table~\ref{tab:sla-post} show SLA violations for POST requests.

\begin{table}[t]
    \caption{SLA violation counts for POST requests}\label{tab:sla-post}
    \centering
    \small
    \begin{tabular}{lccc}
        \toprule
        \textbf{Algorithm} & \textbf{Flexible} & \textbf{Moderate} & \textbf{Strict}\\
        \midrule
        Total Requests & 2,550,000 & 1,220,000 & 1,220,000\\
        \midrule
        Default & 92,055 & 150,060 & 273,036\\
        THPA & 14,280 & 116,876 & 229,360\\
        PPA & 15,555 & 88,084 & 121,268\\
        \textbf{Hybrid} & \textbf{3,315} & \textbf{40,748} & \textbf{66,002}\\
        \bottomrule
    \end{tabular}
\end{table}

For the flexible category, Default was worst (3.61\%, including dropped requests). Reactive and proactive performed similarly (0.56\% and 0.61\%)---the lenient threshold means proactive cannot display cold start mitigation benefits. Hybrid achieved 0.13\%, resulting in approximately 99.9\% availability (near ``high availability'').

The moderate threshold proved far more difficult. Default failed for 12.3\% of requests. Proactive demonstrated cold start importance, achieving 7.22\% vs 9.58\% for reactive. Hybrid achieved only 3.34\%.

For the strict threshold, Default was unable to cope (22.38\%). Proactive outperformed reactive (9.94\% vs 18.8\%), clearly showing cold start importance. Hybrid performed significantly better with only 5.41\%.

Over all experiments and thresholds, the hybrid approach served a minimum of 94.5\% of requests in SLA-compliant manner (worst case), and 100\% in best case. The algorithm displayed robustness and adaptability while requiring little to no user customization.

\section{Conclusions and Future Work}
\label{sec:conclusion}

This paper provides a novel, lightweight, and SLA-compliant approach to autoscale resources on a microservice deployed on an edge architecture. The autoscaler architecture is constructed using open source subsystems, implementing a hybrid approach that combines reactive and proactive auto-scaling to address the cold start problem while maintaining simplicity.

The contributions include: (1) identifying major bottlenecks of auto-scaling on edge deployment compared to cloud architecture, (2) designing a novel hybrid auto-scaling architecture specifically for edge paradigms, and (3) streamlining the forecaster to run on resource-limited edge deployments cost-effectively.

The autoscaler was tested on a production-ready social network microservice deployment, and results compared with cutting-edge autoscalers. The proposed autoscaler achieved a maximum SLA violation rate of 5.41\%, compared to 18.8--22.38\% for state-of-the-art autoscalers. Tests further demonstrated that the autoscaler significantly reduced SLA violations while keeping deployment costs low by assigning resources to maintain utilized resources at approximately half of the configured auto-scaling threshold.

Current limitations include single-metric SLA constraints and horizontal-only scaling. Future directions include: multi-variate forecasting (CPU + memory), vertical pod auto-scaling integration, multi-SLA constraint support with weighted importance, and cluster-level auto-scaling for edge node management.

\bibliographystyle{ACM-Reference-Format}
\bibliography{references}

\end{document}